\documentclass[aps,prd,twocolumn,showpacs]{revtex4-1}
\usepackage{mathtools}
\usepackage{amsfonts}
\usepackage{latexsym}
\usepackage{relsize}
\usepackage{euscript}
\usepackage{amssymb}
\usepackage{graphicx}
\usepackage{amsmath}
\usepackage{amsbsy}
\usepackage{amsthm}
\usepackage{bbm}
\usepackage{bm}
\usepackage{epsfig}
\usepackage{epstopdf}
\usepackage{dsfont}
\usepackage{color}
\usepackage[colorlinks]{hyperref}
\usepackage[figure,table]{hypcap}
\usepackage[symbol]{footmisc}
\usepackage{enumerate}
\usepackage{geometry}
\hypersetup{
	bookmarksnumbered,
	pdfstartview={FitH},
	citecolor={darkgreen},
	linkcolor={darkred},
       linktoc={page},
	urlcolor={darkblue},
	pdfpagemode={UseOutlines}}
\definecolor{darkgreen}{RGB}{50,190,50}
\definecolor{darkblue}{RGB}{0,0,190}
\definecolor{darkred}{RGB}{238,0,0}
\usepackage{soul}

\numberwithin{equation}{section}

\newcommand{\BbbR}{\mathbb{R}}
\newcommand{\BbbZ}{\mathbb{Z}}

\DeclareMathOperator{\sgn}{sgn}

\newcommand{\mralpha}{{{}_{\text{o}}\alpha}}
\newcommand{\mrbeta}{{{}_{\text{o}}\beta}}

\newcommand{\smoothalpha}{{{}_{\text{s}}\alpha}}
\newcommand{\smoothbeta}{{{}_{\text{s}}\beta}}

\newcommand{\mrA}{{{}_{\text{o}}A}}

\newcommand{\smoothA}{{{}_{\text{s}}A}}

\newcommand{\baremass}{{\mu}}

\newcommand{\kperp}{k_{\perp}}

\newcommand{\ztilde}{{\tilde{z}}}

\begin{document}
\title[Scalar, spinor, and photon fields under relativistic cavity motion]
{Scalar, spinor, and photon fields under relativistic cavity motion}

\author{Nicolai Friis}
\email{pmxnf@nottingham.ac.uk}
\author{Antony R. Lee}
\email{pmxal3@nottingham.ac.uk}
\author{Jorma Louko}
\email{jorma.louko@nottingham.ac.uk}
\affiliation{
School of Mathematical Sciences,
University of Nottingham,
University Park,
Nottingham NG7 2RD,
United Kingdom}


\begin{abstract}
We analyse quantised scalar, spinor, and photon fields in a
mechanically rigid cavity that is accelerated in Minkowski spacetime,
in a recently introduced perturbative small-acceleration formalism
that allows the velocities to become relativistic, with a view to
applications in relativistic quantum information. A~scalar field is analysed with both
Dirichlet and Neumann boundary conditions,
and a photon field under perfect conductor boundary conditions
is shown to decompose into Dirichlet-like
and Neumann-like polarisation modes.
The Dirac spinor is analysed with
a nonvanishing mass and with dimensions
transverse to the acceleration,
and the MIT bag boundary condition is shown to exclude zero modes.
Unitarity of time evolution holds for smooth accelerations but fails for
discontinuous accelerations in spacetime dimensions
$(3+1)$ and higher.
As an application, the experimental desktop mode-mixing
scenario proposed for a scalar field by
Bruschi \textit{et al.} [\href{http://dx.doi.org/10.1088/1367-2630/15/7/073052}{New\ J.}\ \href{http://dx.doi.org/10.1088/1367-2630/15/7/073052}{Phys.\ \textbf{15}, 073052 (2013)}]
is shown to apply also to the photon field.
\end{abstract}

\date{September 2013}


\maketitle




\section{Introduction}
\label{sec:intro}

A relativistic quantum field is affected by the kinematics of the
spacetime in which the field lives. Well-known examples are the
Hawking and Unruh effects, associated with black holes and accelerated
observers \cite{hawking,unruh,Crispino:2007eb}, the dynamical (or
nonstationary) Casimir effect
(DCE)~\cite{moore-dyncas,Reynaud1,Dodonov:advchemphys,Dodonov:2010zza},
associated with moving boundaries, and cosmological particle
creation~\cite{parker-I,parker-II}. Similar effects have been
predicted to occur in condensed matter laboratory systems, where the
prospects of experimental verification may be significantly
better~\cite{Barcelo:2005fc}. The effects could be potentially
harnessed to serve quantum information tasks, with current
and near-foreseeable technology, including quantum communication
between satellites~\cite{Rideout:2012jb}.

In this paper we consider a quantum field confined in a cavity that
moves in Minkowski spacetime. The cavity is assumed to be mechanically
rigid, as seen in its instantaneous rest frame, and the acceleration
is assumed to be small in magnitude, compared with the inverse linear
dimensions of the cavity. Under these assumptions the evolution of a
scalar field in the cavity can be solved in a recently developed
formalism that treats the
acceleration perturbatively but allows the velocities, the travel
times, and the travel distances to remain arbitrary, and in particular
allows the velocities to become
relativistic~\cite{Bruschi:2011ug,Bruschi:2012pd,Bruschi:2013vk}. For
acceleration with constant direction, the notion of a relativistic
rigid body can be implemented to all orders in the perturbative
expansion, and for acceleration with varying direction, the formalism
has been developed to first order in the acceleration without
relativistic ambiguities~\cite{Bruschi:2012pd}.

While this small acceleration formalism overlaps in part with
situations covered by the small distance approximations often
considered in the DCE literature
\cite{Dodonov:advchemphys,Dodonov:2010zza}, and by other
approximation schemes~\cite{DalvitMazzitelli1999,AlvesGranhenPires2010},
its novelty is in the ability to
accommodate relativistic velocities in a systematic fashion.
Applications to
quantum information tasks in relativistic or potentially relativistic
contexts have been analysed in
\cite{Bruschi:2011ug,Bruschi:2012pd,Bruschi:2013vk,Friis:2011yd,Friis:2012tb,Bruschi:2012uf,Alsing:Fuentes:2012,Friis:2012ki,Friis:2012nb,Friis:2012cx}.
In particular, the formalism is applicable to a cavity whose motion is
implemented by
superconducting quantum interference device (SQUID) circuits
without mechanically moving
parts~\cite{Friis:2012cx}.
A~generalisation to massless fermions in a
$(1+1)$-dimensional cavity is given in~\cite{Friis:2011yd}.

The main purpose of this paper is to adapt the analysis of a scalar field
in the cavity to the electromagnetic field,
with perfect conductor boundary conditions at the cavity walls.
The interest of this question arises from the
traditional prime suspect role of the
electromagnetic field in experimental scenarios that involve
acceleration effects
\cite{moore-dyncas,Reynaud1,Dodonov:advchemphys,Dodonov:2010zza},
including the recent experiments in which acceleration is simulated by
SQUID circuits~\cite{wilson-etal}.  We find that the electromagnetic
field decomposes into two sets of polarisation modes, one similar to a
Dirichlet scalar field and the other similar to a Neumann scalar
field.  The results for the evolution of the electromagnetic field
hence follow in a straightforward fashion from those for the Dirichlet
scalar field, found in \cite{Bruschi:2011ug,Bruschi:2012pd}, and those
for the Neumann scalar field, which we provide in this paper.  In
particular, our results confirm that the experimental scenario of a
cavity accelerated on a desktop,
proposed and analysed for a scalar field in~\cite{Bruschi:2012pd},
applies also to the photon field.

A second purpose is to address a Dirac spinor at a
generality that covers a $(3+1)$-dimensional cavity, generalising the
case of a massless $(1+1)$ field analysed
in~\cite{Friis:2011yd}. This question is motivated by the prospect of
simulating acceleration effects for fermions in solid state analogue
systems~\cite{boada:dirac,ZhangWangZhu2012,Iorio2012}.
After finding the general family of boundary
conditions that ensures a vanishing probability current through
the walls, we specialise to the MIT bag boundary
condition~\cite{Chodos:1974je,Elizalde:1997hx}, which arises when the
cavity field is matched to a highly massive field in the exterior
and the exterior mass is taken to infinity. We find that
the MIT bag boundary condition leads to a charge conjugation
symmetric spectrum without zero modes, independently of the field mass
or of effects from dimensions transverse to the acceleration.  We also
present the explicit Fourier transform formulas for the Bogoliubov
coefficients in the case when the acceleration varies smoothly in time,
generalising the scalar field formulas given in~\cite{Bruschi:2012pd}.
We use the insights gathered about the fermionic Bogoliubov transformations
to comment on the implications for quantum information purposes, such as
discussed in~\cite{Friis:2011yd,Friis:2012tb,Friis:2012ki}.

A third purpose is to examine whether the time evolution
of the cavity field is implementable as a unitary transformation in
the Fock space. (We thank Pablo Barberis-Blostein and Ivette Fuentes
for drawing our attention to this question.) As the potential failure
of unitarity is governed by the deep ultraviolet regime of the theory,
the issue here is whether
any predictions computed from the formalism are sensitive to the
idealisations made in the ultraviolet.
Comfortingly, we find that the evolution is unitary whenever
the acceleration varies smoothly in time.
In the limit of discontinuous acceleration, unitarity however fails
in spacetime dimensions $(3+1)$ and higher.

A fourth purpose is to give a proper justification to certain
technical properties that have been stated and utilised in earlier
papers~\cite{Bruschi:2011ug,Bruschi:2012pd,Bruschi:2013vk,Friis:2011yd,Friis:2012tb,Bruschi:2012uf,Alsing:Fuentes:2012,Friis:2012ki,Friis:2012nb,Friis:2012cx}.
In particular, we explain how the direction of the acceleration comes
to be encoded in the Bogoliubov coefficient formulas.

We begin in Section \ref{sec:scalar-Dirichlet}
by recalling the Dirichlet scalar field
analysis that was outlined in~\cite{Bruschi:2011ug},
establishing the notation for the rest of the paper.
The Neumann scalar field is
addressed in Sec.~\ref{sec:scalar-Neumann}.
The electromagnetic field is addressed in Secs.~\ref{sec:Maxwell-static-cavity}
and
\ref{sec:Maxwell-Rindler-cavity},
and the Dirac field in
Sec.~\ref{sec:Diracspinor}.
Unitarity of the evolution is analysed in Sec.~\ref{sec:unitarity}, with
auxiliary asymptotic estimates deferred to the Appendix.
The results are summarised and discussed in Sec.~\ref{sec:conclusions}.

Our metric signature is mostly plus, and we use units in which
$c=\hbar=1$.


\section{$(1+1)$ Dirichlet scalar field}
\label{sec:scalar-Dirichlet}

In this section we address a real scalar field of strictly positive
mass in $(1+1)$-dimensional Minkowski spacetime, with Dirichlet
boundary conditions at the cavity walls. While the core results can
be found in earlier short format
papers~\cite{Bruschi:2011ug,Bruschi:2012pd,Bruschi:2013vk,Bruschi:2012uf},
our purpose here is to be sufficiently self-contained to allow a
direct comparison to the Maxwell field analysis in
Sec.~\ref{sec:Maxwell-Rindler-cavity}.

\subsection{Inertial cavity}
\label{subsec:dirichlet-inertial}

Let $\phi$ be a real scalar field of mass~$\baremass>0$ in
$(1+1)$-dimensional Minkowski spacetime,
satisfying the Klein-Gordon equation
\begin{align}
\left(-\square + \baremass^2\right) \phi =0\ ,
\end{align}
where $\square$ is the scalar Laplacian.
The field is confined in a cavity that may move but maintains
a prescribed length $L>0$ in its instantaneous rest frame.
The field is assumed to satisfy Dirichlet boundary conditions at the cavity walls.

When the cavity is inertial,
we may introduce Minkowski coordinates $(t,z)$ in which the
metric reads
\begin{align}
ds^2 = - dt^2 + dz^2 \ ,
\label{eq:1+1Minkowski}
\end{align}
and the walls are, respectively, at
$z=z_0$ and $z=z_1 := z_0+L$,
dragged along the timelike Killing vector $\partial_t$.
$z_0$ could be set to zero without loss of generality,
but leaving $z_0$ unspecified for the moment will be useful for
matching to accelerated motion below.

The Klein-Gordon inner product takes the form
\begin{align}
(\phi_1,\phi_2) = -i \int_{z_0}^{z_1}
\phi_1 \overleftrightarrow{\partial_t} \> \overline{\phi_2}
\, dz
\ ,
\label{eq:KGip-Minkowski}
\end{align}
where the overline denotes complex conjugation
(we adopt the conventions of~\cite{byd} in which the
inner product is antilinear in the second argument).
A standard basis of field modes that are of positive frequency with respect to
$\partial_t$ and orthonormal in the Klein-Gordon
inner product \eqref{eq:KGip-Minkowski} is
\begin{subequations}
\label{MinkowskiSolutions-all}
\begin{align}
\phi^{M}_n (t,z) &:=
\frac{1}{\sqrt{\omega_{n}L}}
\sin \! \left(\frac{n\pi (z-z_0)}{L}\right)
e^{-i\omega_{n}t}
\ ,
\label{MinkowskiSolutions-function}
\\[1ex]
\omega_{n} &:= \sqrt{\baremass^2 + {(\pi n/L)}^2}
\ ,
\label{eq:inertial-omega-def}
\end{align}
\end{subequations}
where $n=1, 2, \ldots$.
The phase in \eqref{MinkowskiSolutions-function}
has been chosen so that $\partial_z \phi^{M}_n |_{z=z_0}>0$ at $t=0$.


\subsection{Uniformly accelerated cavity}
\label{subsec:dirichlet-accelerated}

When the cavity is uniformly accelerated,
in the sense of being dragged along a boost Killing vector,
we may introduce Rindler coordinates $(\eta,\chi)$
\cite{takagi} in which
\begin{align}
ds^2 = - \chi^2 \, d\eta^2 + d\chi^2 \ ,
\label{eq:1+1Rindlermetric}
\end{align}
with $-\infty<\eta<\infty$ and $0<\chi<\infty$,
and the cavity walls are, respectively, at
$\chi=\chi_0>0$ and $\chi=\chi_1 := \chi_0+L$.
The boost Killing vector is $\partial_\eta$.
It is convenient to parametrise the geometry of the accelerated cavity
by the pair $(h,L)$, where the dimensionless parameter $h$
lies in the interval $0<h<2$,
such that
\begin{subequations}
\label{eq:h-introd}
\begin{align}
\chi_0 &= \left(\frac{1}{h} - \frac12 \right) \! L
\ ,
\\
\chi_1 &= \left(\frac{1}{h} + \frac12 \right) \! L
\ .
\end{align}
\end{subequations}
The proper acceleration at the centre of the cavity,
at $\chi = (\chi_0+\chi_1)/2$,
equals~$h/L$.
Note that the proper acceleration is not uniform within the cavity:
each worldline of constant $\chi$ has proper acceleration $1/\chi$,
and the proper accelerations at the cavity walls are hence,
respectively, $1/\chi_0$ and $1/\chi_1$.
The upper bound on $h$ comes from the condition that
the proper acceleration at both cavity walls remain finite.

The Klein-Gordon inner product takes the form
\begin{align}
(\phi_1,\phi_2) = -i \int_{\chi_0}^{\chi_1}
\phi_1 \overleftrightarrow{\partial_\eta} \> \overline{\phi_2}
\, \chi^{-1}\, d\chi
\ .
\label{eq:KGip-Rindler}
\end{align}
By separation of variables~\cite{takagi}, we find that
a basis of field field modes that are of positive frequency with respect to
$\partial_\eta$ and orthonormal in the Klein-Gordon
inner product~\eqref{eq:KGip-Rindler}~is
\begin{subequations}
\label{RindlerSolutions-function}
\begin{align}
 \hspace{13mm} &\begin{aligned}
    \mathllap{\phi^{R}_n (\eta,\chi)} &\phantom{:}= f_{n}(\chi)e^{-i\Omega_n \eta}\ ,
  \end{aligned}\\
  &\begin{aligned}
    \mathllap{f_{n}(\chi)} &:= N_n \big[ I_{-i\Omega_n}(\baremass \chi_0) I_{i\Omega_n}(\baremass \chi)\\
      &\hspace{15mm} - I_{i\Omega_n}(\baremass \chi_0) I_{-i\Omega_n}(\baremass \chi) \big] \ ,
  \end{aligned}
\end{align}
\end{subequations}
where $n=1,2,\ldots$, $I$ is the modified Bessel function of the first
kind~\cite{nist-dig-library}, the eigenfrequencies $\Omega_n$ are
determined by the boundary condition $\phi^{R}_n (\eta,\chi_1)=0$ and
are ordered so that $0 < \Omega_1 \le \Omega_2 \le \cdots$, and $N_n$
is a normalisation constant.  We shall return to the phase choice of
$N_n$ in subsection~\ref{subsec:dirichlet-matching}.


Note that both $\eta$ and $\Omega_n$ are dimensionless.
As the proper time at the centre of the cavity equals $L\eta/h$,
the angular frequency of $\phi^{R}_n$ with respect to this proper time is $h\Omega_n/L$.


\subsection{Matching}
\label{subsec:dirichlet-matching}

Consider now a cavity whose motion turns instantaneously from inertial
to uniform acceleration, so that the wall velocities are
continuous but the proper accelerations have a finite discontinuity.
We take the inertial segment to be as in subsection
\ref{subsec:dirichlet-inertial} for $t\le0$ and the uniformly
accelerated segment to be as in subsection
\ref{subsec:dirichlet-accelerated} for $\eta\ge0$.

\begin{figure}
\includegraphics[width=0.45\textwidth]{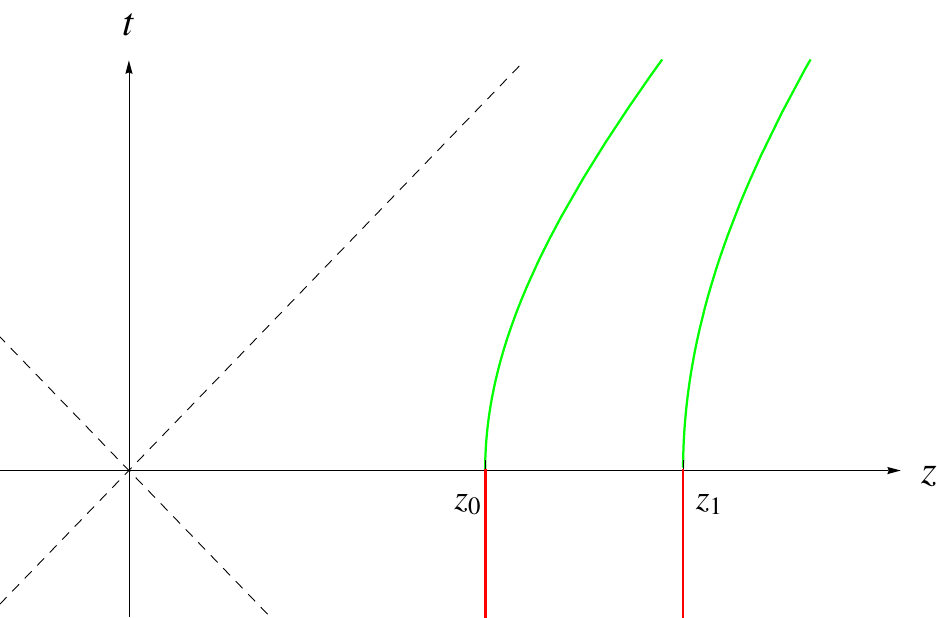}
\caption{\label{fig:mink-to-rindler}Matching an inertial cavity to a
uniformly accelerating cavity in $(1+1)$-dimensional Minkowski
space, in the global Minkowski coordinates $(t,z)$. For $t\le0$ the
cavity is inertial, following the orbits of the time translation
Killing vector~$\partial_t$: the world lines of the walls are, respectively,
$z=z_0$ and $z=z_1$, where $0<z_0<z_1$. For $t\ge0$ the cavity is
uniformly accelerated towards increasing~$z$,
in the sense that it follows the orbits of the
boost Killing vector $z\partial_t +
t\partial_z$: the world lines of the walls are, respectively,
$z = \sqrt{z_0^2+t^2}$ and $z = \sqrt{z_1^2+t^2}$.}
\end{figure}

To begin with, suppose that the acceleration is towards
increasing~$z$.  The transformation relating the Minkowski and Rindler
coordinates is then \cite{takagi}
\begin{subequations}
\label{eq:Rindlertransformation}
\begin{align}
t &= \chi \sinh\eta
\ ,
\\
z &= \chi \cosh\eta
\ ,
\end{align}
\end{subequations}
and the cavity wall loci at $t=0$ in the two coordinate systems are related
by $z_0 = \chi_0$ and $z_1 = \chi_1$,
as shown in Fig.~\ref{fig:mink-to-rindler}.

We write the Bogoliubov transformation from the
Minkowski modes to the Rindler modes as
\begin{align}
\phi^{R}_m = \sum_n \, \left(
\mralpha_{mn} \phi^{M}_n + \mrbeta_{mn} \overline{\phi^{M}_n}
\right)
\ .
\label{eq:dirichlet-bogo}
\end{align}
From \eqref{eq:dirichlet-bogo} and the orthonormality of the Minkowski modes,
we have \cite{byd}
\begin{subequations}
\label{eq:alphabetaipformulas}
\begin{align}
\mralpha_{mn}
& =
\bigl(\phi^{R}_m,\phi^{M}_n\bigr)
\ ,
\\
\mrbeta_{mn}
& =
- \bigl(\phi^{R}_m,\overline{\phi^{M}_n} \, \bigr)
\ ,
\end{align}
\end{subequations}
where the inner products may be evaluated by \eqref{eq:KGip-Minkowski}
at $t=0$ or equivalently by \eqref{eq:KGip-Rindler} at $\eta=0$.
While these inner products
do not appear to have expressions in terms of known functions, they
can be given perturbative small $h$ expansions~\cite{Bruschi:2011ug}.
As small $h$ means small acceleration, in the leading order
$\phi^{R}_n$
must be equal to $\phi^{M}_n$ up to a phase factor, and we fix this
factor to unity by choosing the phase of $N_n$ in
\eqref{RindlerSolutions-function} so that $\partial_\chi \phi^{R}_n
|_{\chi=\chi_0}>0$ at $\eta=0$.  The sub-leading terms in $\phi^{R}_n$
can then be written as a power series in~$h$, with the help of uniform
asymptotic expansions of the modified Bessel functions in
\eqref{RindlerSolutions-function}~\cite{nist-dig-library,dunster:1990:bfp}.
We find that $(h\Omega_n)/(L\omega_n) = 1 + O(h^2)$, and the
expressions for the Bogoliubov coefficients to linear order in $h$ are
given in equations (7) in \cite{Bruschi:2011ug} and can be rearranged
to read
\begin{subequations}
\label{eq:dirichlet-bogocoeffs-linear}
\begin{align}
 &\begin{aligned}
    \mralpha_{nn} &= 1 + O(h^2) \ ,
  \end{aligned}\\
  &\begin{aligned}
    \mralpha_{mn} &= \frac{\pi^2 m n \bigl(-1+{(-1)}^{m+n}\bigr)}
{L^4\left(\omega_m - \omega_n\right)^3
\sqrt{\omega_m \omega_n}}
\, h + O(h^2)\\
&\hspace{35mm}\hbox{(for $m\ne n$)}
  \end{aligned} \ ,\\
  &\begin{aligned}
    \mrbeta_{mn} &= \frac{\pi^2 m n \bigl(1 - {(-1)}^{m+n}\bigr)}
{L^4\left(\omega_m + \omega_n\right)^3
\sqrt{\omega_m \omega_n}}
\, h + O(h^2) \ .
\label{eq:dirichlet-betacoeffs-linear}
  \end{aligned}
\end{align}
\end{subequations}
The expansion \eqref{eq:dirichlet-bogocoeffs-linear} holds as $h\to0$
for fixed~$m$ and~$n$, but the size of the error terms depends on~$m$
and~$n$, and the expansion is hence not uniform in the indices of the
Bogoliubov coefficients. It can however be verified
\cite{Bruschi:2011ug} that when the $h^2$ terms are included
in~\eqref{eq:dirichlet-bogocoeffs-linear}, these expansions satisfy
the Bogoliubov identities \cite{byd} perturbatively to order~$h^2$,
which provides an internal consistency check on the perturbative
formalism. (We note in passing that formula (7a) in
\cite{Bruschi:2011ug} contains a typographic error in that the $h^2$
contribution to $\mralpha_{nn}$ given therein should contain the
additional term $+\frac{7}{16}\frac{M^4}{\pi^6n^6}h^2$.)

Finally, recall that above we have assumed the acceleration to be
towards increasing~$z$. For acceleration towards decreasing~$z$, we
may proceed similarly, introducing the leftward Rindler coordinates
$(\tilde\eta,\tilde\chi)$ by
\begin{subequations}
\label{eq:leftRindlertransformation}
\begin{align}
t &= \tilde\chi \sinh\tilde\eta
\ ,
\\
z &= - \tilde\chi \cosh\tilde\eta
\ ,
\end{align}
\end{subequations}
in which the metric is as in \eqref{eq:1+1Rindlermetric} but with tildes.
The loci of the cavity walls at $t=0$ are now related by
$\tilde\chi_0 = - z_1$ and
$\tilde\chi_1 = - z_0 = - z_1 + L$, where $z_1<0$.
The only difference in the analysis is that the phases
of the new Rindler modes must still be matched to those of the
Minkowski modes $\phi^{M}_n$~\eqref{MinkowskiSolutions-function},
which were already fixed above. Since a left-right reflection changes
$\phi^{M}_n$ \eqref{MinkowskiSolutions-function}
by the factor~${(-1)}^{n+1}$, the formulas for $\mralpha_{mn}$ and
$\mrbeta_{mn}$ in leftward acceleration are obtained from those in
rightward acceleration by keeping $h$ positive and inserting the phase
factors ${(-1)}^{m+n}$. To linear order in~$h$, this can be
implemented by taking the formulas
\eqref{eq:dirichlet-bogocoeffs-linear} to hold for both signs of~$h$,
with positive (respectively, negative) $h$ denoting acceleration
towards increasing (decreasing)~$z$.  We have verified that this
implementation holds also when the $h^2$ contributions are included
in~\eqref{eq:dirichlet-bogocoeffs-linear}.

\subsection{Time-dependent acceleration}
\label{subsec:dirichlet-continuous}

For cavity motion in which the acceleration is piecewise constant in
time, we can compose inertial and uniformly accelerated segments by
the above Minkowski-to-Rindler transformation and its
inverse~\cite{Bruschi:2011ug}. For motion in which the acceleration
is not necessarily piecewise constant in time, we can pass to the
limit in which the constant acceleration segments have vanishing
duration~\cite{Bruschi:2012pd}.


To establish the notation, let $\tau$ denote the proper time and
$a(\tau)$ the proper acceleration at the centre of the cavity, such
that positive (negative) $a(\tau)$ means acceleration towards
increasing (decreasing) $z$ in the global Minkowski coordinates. Let
the acceleration vanish in the initial inertial region $\tau\le\tau_0$
and in the final inertial region $\tau\ge \tau_f$.  To linear order in
the acceleration, the Bogoliubov coefficient matrices
$(\smoothalpha,\smoothbeta)$ between the initial and final inertial
regions have then the expressions \cite{Bruschi:2012pd}
\begin{subequations}
\label{eq:ABhat-int-components}
\begin{align}
  &\begin{aligned}
    \smoothalpha_{nn} &= e^{i{\omega}_n (\tau_f-\tau_0)} \ ,
  \end{aligned}\\
  &\begin{aligned}
    \smoothalpha_{mn} &= i L (\omega_m-\omega_n) \hat\alpha_{mn}(M) \, e^{i{\omega}_m (\tau_f-\tau_0)}\\
	&\hspace{5mm}\times\int_{\tau_0}^{\tau_f} e^{-i({\omega}_m - {\omega}_n) (\tau-\tau_0)} \, a(\tau) \, d\tau \\
	&\ \hspace{35mm} \hbox{(for $m\ne n$)},
  \end{aligned}\\
  &\begin{aligned}
    \smoothbeta_{mn} &= i L (\omega_m+\omega_n) \hat\beta_{mn}(M) \, e^{i{\omega}_m (\tau_f-\tau_0)}\\
	&\hspace{5mm}\times\int_{\tau_0}^{\tau_f} e^{-i({\omega}_m + {\omega}_n) (\tau-\tau_0)} \, a(\tau) \, d\tau \ ,
  \end{aligned}
\label{eq:Bhat-int-components}
\end{align}
\end{subequations}
where $\hat\alpha_{mn}(M)$ and $\hat\beta_{mn}(M)$ are the
coefficients of $h$ in the expansions
\eqref{eq:dirichlet-bogocoeffs-linear} of $\mralpha_{mn}$
and~$\mrbeta_{mn}$, and we have indicated explicitly that these
coefficients depend on $\baremass$ and $L$ only through the
dimensionless combination $M := \baremass L$.  To linear order in the
acceleration, the Bogoliubov coefficients are hence obtained by
Fourier transforming the acceleration.

\section{$(1+1)$ Neumann scalar field}
\label{sec:scalar-Neumann}

In this section we adapt the analysis of Section \ref{sec:scalar-Dirichlet}
to a scalar field with Neumann boundary conditions at the cavity walls.
To avoid cluttering the notation, we shall suppress
in the field modes and the Bogoliubov coefficients
an explicit index that would distinguish the Dirichlet and Neumann boundary conditions.

For the inertial cavity, a standard basis of field modes that are of
positive frequency with respect to
$\partial_t$ and orthonormal in the Klein-Gordon
inner product \eqref{eq:KGip-Minkowski} is
\begin{align}
\!\!\phi^{M}_n (t,z) &:=
\begin{cases}
\displaystyle
\frac{1}{\sqrt{2\omega_0 L}}
\,
e^{-i\omega_{0}t}\hspace{5mm}(n=0)
\ ,
\\[3ex]
\displaystyle
\frac{1}{\sqrt{\omega_{n}L}}
\cos \! \left[\frac{n\pi (z-z_0)}{L}\right]\!\!
e^{-i\omega_{n}t},
\\
\hspace{20mm} (n=1,2,\ldots )
\end{cases}
\label{MinkowskiSolutions-function-N}
\end{align}
where $n=0,1,2, \ldots$ and
$\omega_{n}$ is given by \eqref{eq:inertial-omega-def}.
The phase has been chosen so that $\phi^{M}_n |_{z=z_0}>0$ at $t=0$.

For the uniformly accelerated cavity, a basis
of field modes that are of
positive frequency with respect to
$\partial_\eta$ and orthonormal in the Klein-Gordon
inner product \eqref{eq:KGip-Minkowski} is
\begin{subequations}
\label{RindlerSolutions-function-N}
\begin{align}
  \phantom{i + j + k}
  &\begin{aligned}
    \mathllap{\phi^{R}_n (\eta,\chi)} &\phantom{:}= f_{n}(\chi)e^{-i\Omega_n \eta} \ ,
  \end{aligned}\\
  &\begin{aligned}
    \mathllap{f_{n}(\chi)} &:= N_n\bigl[I'_{-i\Omega_n}(\baremass \chi_0) I_{i\Omega_n}(\baremass \chi)\\
      &\hspace{15mm} - I'_{i\Omega_n}(\baremass \chi_0) I_{-i\Omega_n}(\baremass \chi)\bigr] \ ,
  \end{aligned}
\end{align}
\end{subequations}
where $n=0,1,2,\ldots$,
the prime denotes derivative with respect to the argument,
the eigenfrequencies $\Omega_n$
are determined by the boundary condition $\partial_\chi\phi^{R}_n |_{\chi=\chi_1} =0$
and are ordered so that $0 < \Omega_0 \le \Omega_1 \le \cdots$,
and $N_n$ is a normalisation constant.
The angular frequency of $\phi^{R}_n$ with respect to the proper
time at the centre of the cavity is $h\Omega_n/L$.

Matching the inertial segment at $t\le0$ to a uniformly accelerated
segment at $\eta\ge0$ is done as in
subsection~\ref{subsec:dirichlet-matching}.  When the acceleration is
towards increasing~$z$, we relate the Minkowski and Rindler
coordinates by \eqref{eq:Rindlertransformation} and choose the phase
of the normalisation constant $N_n$ so that $\phi^{R}_n (0,\chi_0)>0$.
We again find that $(h\Omega_n)/(L\omega_n) = 1 + O(h^2)$,
and the expressions for the Bogoliubov coefficients to linear order in
$h$ read
\begin{widetext}
\begin{subequations}
\label{eq:neumann-bogocoeffs-linear}
\begin{align}
\mralpha_{nn} &= 1 + O(h^2)
\ ,
\\[1ex]
\mralpha_{mn} &=
\begin{dcases*}
\displaystyle
\frac{\left(\omega_m \omega_n - \baremass^2 \right) \bigl(-1+{(-1)}^{m+n}\bigr)}
{L^2\left(\omega_m - \omega_n\right)^3
\sqrt{\omega_m \omega_n}}
\, h
\hspace{-3mm} & $+ O(h^2)$ for $m>0$, $n>0$ and $m\ne n$,
\\
\frac{\left(\omega_m \omega_n - \baremass^2 \right) \bigl(-1+{(-1)}^{m+n}\bigr)}
{\sqrt{2} \, L^2\left(\omega_m - \omega_n\right)^3
\sqrt{\omega_m \omega_n}}
\, h
\hspace{-3mm} & $+ O(h^2)$ for $m>n=0$ or $n>m=0$,
\end{dcases*}
\end{align}
\begin{align}
\mrbeta_{mn} &=
\begin{dcases*}
\displaystyle
\frac{\left(\omega_m \omega_n + \baremass^2 \right)
\bigl(1 - {(-1)}^{m+n}\bigr)}
{L^2\left(\omega_m + \omega_n\right)^3
\sqrt{\omega_m \omega_n}}
\, h
& \hspace{-3mm}$+ O(h^2)$ for $m>0$ and $n>0$,
\\
\frac{\left(\omega_m \omega_n + \baremass^2 \right)
\bigl(1 - {(-1)}^{m+n}\bigr)}
{\sqrt{2} \, L^2\left(\omega_m + \omega_n\right)^3
\sqrt{\omega_m \omega_n}}
\, h
& \hspace{-3mm}$+ O(h^2)$ for $m>n=0$ or $n>m=0$.
\end{dcases*}
\label{eq:neumann-betacoeffs-linear}
\end{align}
\end{subequations}
\end{widetext}
As with the Dirichlet boundary condition, the small $h$ expansion is
not uniform in the indices of the Bogoliubov coefficients, but we have
again verified that when the $h^2$ terms are included
in~\eqref{eq:neumann-bogocoeffs-linear}, these expansions satisfy the
Bogoliubov identities \cite{byd} perturbatively to order~$h^2$, which
provides an internal consistency check on the formalism.

To accommodate both directions of acceleration,
we proceed as with the Dirichlet boundary conditions.
Taking positive (respectively, negative) $h$ to denote acceleration
towards increasing (decreasing)~$z$, we find that the formulas
\eqref{eq:neumann-bogocoeffs-linear}
hold for both signs of~$h$, and they continue to hold for both signs
of $h$ also when the $h^2$ terms are included.

Finally, cavity motion with time-dependent acceleration can be handled
as with the Dirichlet conditions. To linear order in the acceleration,
the Bogoliubov coefficient matrices $(\smoothalpha,\smoothbeta)$
between initial and final inertial regions are given
by~\eqref{eq:ABhat-int-components}, where $\hat\alpha_{mn}(M)$ and
$\hat\beta_{mn}(M)$ are now the coefficients of $h$ in the
expansions~\eqref{eq:neumann-bogocoeffs-linear}, and we have indicated
explicitly that these coefficients depend on $\baremass$ and $L$ only
through the dimensionless combination $M := \baremass L$.

\section{Curved spacetime Maxwell field in a static perfect conductor cavity}
\label{sec:Maxwell-static-cavity}

In this section we write down the action of the Maxwell field in a
$(3+1)$-dimensional static but possibly curved spacetime, in a static
cavity with perfect conductor boundary conditions.  The main issue is
to adapt the gauge choice both to the staticity \cite{Zhao:2011nq} and
to the boundary conditions~\cite{jackson-bible}.

\subsection{Gauge choice}

We consider a static $(3+1)$-dimensional spacetime, working in coordinates
$(t,x^1,x^2,x^3)$ in which the metric reads
\begin{align}
ds^2 = - N^2 dt^2 + h_{ij} \, dx^i \, dx^j
\ ,
\label{eq:staticmetric}
\end{align}
where the Latin indices $i,j,\ldots$ from the
middle of the alphabet take values in $\{1,2,3\}$,
$N>0$, $h_{ij}$ is positive definite, and neither $N$ nor $h_{ij}$ depends on~$t$.
The timelike hypersurface-orthogonal Killing vector is~$\partial_t$,
and it is orthogonal to the hypersurfaces of constant~$t$.
We postpone issues of spatial boundary conditions to
subsection~\ref{subsec:Maxwell-cavity-conditions}.

The Maxwell action reads
\begin{align}
S = - \frac14 \int d^4x \sqrt{-g} \, F_{ab} F^{ab}
\ ,
\label{eq:Maxwell-action}
\end{align}
where
$F_{ab} = \partial_a A_b - \partial_b A_a$,
$A_a$ is the electromagnetic potential,
the spacetime indices $a,b,\ldots$ are raised and
lowered with the metric $g_{ab}$~\eqref{eq:staticmetric}
and $g = \det(g_{ab})$.
Following Dirac's
procedure~\cite{dirac-yeshiva,henneaux-teitelboim:book},
the action can be put in the Hamiltonian form
\begin{align}
&S = \int dt \, d^3x
\, \Bigl(
\pi^i \dot A_i
+ A_0 \partial_i \pi^i
- \frac{N}{2 \sqrt{h}} \, \pi_i \pi^i
\notag
\\
&\hspace{16ex}
- \frac14 N \sqrt{h} \, F_{ij} F^{ij}
\Bigr)
\ ,
\label{eq:Maxwell-Hamiltonian-action}
\end{align}
where $F_{ij} = \partial_i A_j - \partial_j A_i$,
the overdot denotes derivative with respect to~$t$,
the spatial indices are raised and lowered with $h_{ij}$ and its inverse
$h^{ij}$, and $h = \det(h_{ij})$.

Variation of \eqref{eq:Maxwell-Hamiltonian-action}
with respect to $\pi^i$
and $A_i$ gives the dynamical field equations
\begin{subequations}
\label{eq:Maxwell-dyn-eom}
\begin{align}
\dot A_i &= \frac{N}{\sqrt{h}}\pi_i
+ \partial_i A_0
\label{eq:Maxwell-dyn-eom-Adot}
\ ,
\\
\dot \pi^i &=
\partial_j
\! \left(
N \sqrt{h} \, F^{ji}
\right)
\notag
\\
&=
\sqrt{h} \, \nabla_j
\! \left(
N F^{ji} \right)
\ ,
\label{eq:Maxwell-dyn-eom-pidot}
\end{align}
\end{subequations}
where $\nabla$ denotes the covariant derivative with respect to~$h_{ij}$.
Variation with respect to $A_0$ gives the constraint
\begin{align}
\partial_i \pi^i &=0
\ ,
\label{eq:Maxwell-constraint}
\end{align}
which is preserved in time by~\eqref{eq:Maxwell-dyn-eom}.
In Dirac's terminology, $(A_i, \pi^i)$ is a
canonically conjugate pair of dynamical variables,
while $A_0$ is a Lagrange multiplier
that enforces the first class constraint~\eqref{eq:Maxwell-constraint}.
The Hamiltonian gauge transformations read
\begin{subequations}
\label{eq:ham-gaugetransf}
\begin{align}
\delta A_0 &= \dot \Lambda
\ ,
\label{eq:ham-gaugetransf-multiplier}
\\
\delta A_i &= \partial_i \Lambda
\ ,
\label{eq:ham-gaugetransf-vector}
\\
\delta \pi^i &= 0
\ ,
\end{align}
\end{subequations}
where the function $\Lambda$ is the generator
of the transformation.
These transformations clearly leave the Hamiltonian action
\eqref{eq:Maxwell-Hamiltonian-action} invariant.

We adopt the Coulomb gauge
\begin{subequations}
\label{eq:coulombgauge}
\begin{align}
&\nabla^i \! \left(\frac{A_i}{N}\right) = 0
\ ,
\label{eq:coulombgauge-vector}
\\[1ex]
&A_0 = 0
\ .
\label{eq:coulombgauge-multiplier}
\end{align}
\end{subequations}
The choice \eqref{eq:coulombgauge-vector} can be accomplished on an initial
hypersurface of constant $t$ by the
gauge transformation \eqref{eq:ham-gaugetransf-vector}
by solving an elliptic equation for~$\Lambda$.
The choice \eqref{eq:coulombgauge-multiplier}
for the Lagrange multiplier $A_0$
then preserves
\eqref{eq:coulombgauge-vector} under the time evolution~\eqref{eq:Maxwell-dyn-eom},
using the constraint~\eqref{eq:Maxwell-constraint}.

After the inverse Legendre transform into a Lagrangian formalism
in which $A_i$ satisfies the gauge condition~\eqref{eq:coulombgauge-vector},
the action becomes
\begin{align}
S = \int dt \, d^3x
\left(
\frac{\sqrt{h}}{2 N} \, \dot A_i \dot A^i
- \frac14 N \sqrt{h} \, F_{ij} F^{ij}
\right)
\ .
\label{eq:Maxwell-Lagr-gaugefixed-action}
\end{align}
The field equation reads
\begin{align}
\ddot A_i =
N \nabla^j
\! \left(
N F_{ji} \right)
\ ,
\label{eq:Maxwell-Lagr-gaugefixed-eom}
\end{align}
and the conserved inner product is
\begin{align}
(A_{(1)},A_{(2)}) = -i \int
d^3x \, \frac{\sqrt{h}}{N}\,
\left(A_{(1)}\right)_i \overleftrightarrow{\partial_t} \> \overline{A_{(2)}^i}
\ .
\label{eq:Maxwell-ip}
\end{align}

\subsection{Cavity boundary conditions}
\label{subsec:Maxwell-cavity-conditions}

We consider a cavity whose
walls follow orbits of the Killing vector~$\partial_t$.
The cavity is hence static with respect to~$\partial_t$.

We require $A_i$ to be orthogonal to the
cavity walls.  This implies the conventional perfect conductor
boundary condition that the electric field be orthogonal to the walls
and the magnetic field be parallel to the walls~\cite{jackson-bible}.
This boundary condition annihilates the spatial boundary terms
in the variation of the action
\eqref{eq:Maxwell-Lagr-gaugefixed-action}
so that the equation of
motion \eqref{eq:Maxwell-Lagr-gaugefixed-eom} is obtained. It also annihilates
the boundary terms that arise when the conservation of the
inner product \eqref{eq:Maxwell-ip} is verified. The boundary condition is hence
consistent with the dynamics.

\section{Maxwell field in an accelerated cavity}
\label{sec:Maxwell-Rindler-cavity}

In this section we discuss the Maxwell field in $(3+1)$-dimensional
Minkowski spacetime, in a rigid rectangular cavity that is accelerated
in one of its principal directions. Subsections
\ref{subsec:Maxwell-configuration}--\ref{subsec:Maxwell-secondpol} address the case of uniform acceleration in the gauge-fixed formalism
of Sec.~\ref{sec:Maxwell-static-cavity}. Time-dependent acceleration
is addressed in subsection~\ref{subsec:Maxwell-timedependentacc}.

\subsection{Cavity configuration}
\label{subsec:Maxwell-configuration}

We consider a rectangular cavity with edge lengths $(L_x,L_y,L_z)$, in
uniform acceleration in the $z$ direction. In adapted Rindler
coordinates $(\eta,\chi,x,y)$, the metric reads
\begin{align}
ds^2 = - \chi^2 \, d\eta^2 + d\chi^2 + dx^2 + dy^2
\ ,
\label{eq:3+1Rindlermetric}
\end{align}
and the cavity worldtube is at
\begin{subequations}
\begin{align}
&0 \le x \le L_x
\ ,
\\
&0 \le y \le L_y
\ ,
\\
&\chi_0 \le \chi \le \chi_1
\ ,
\end{align}
\end{subequations}
where $\chi_0>0$ and $\chi_1 = \chi_0 + L_z$. We may parametrise
$\chi_0$ and $\chi_1$ as in \eqref{eq:h-introd} with $L \to L_z$, so
that the dimensionless parameter $h$ satisfies $0<h<2$ and the proper
acceleration at the centre of the cavity equals~$h/L_z$.

We follow the gauge-fixed formalism of Section
\ref{sec:Maxwell-static-cavity} and seek solutions to the field
equation \eqref{eq:Maxwell-Lagr-gaugefixed-eom} with the perfect
conductor boundary conditions by separation of variables. We find that
the field modes that are of positive frequency with respect to
$\partial_\eta$ and orthonormal in the inner product
\eqref{eq:Maxwell-ip} fall into two qualitatively different
polarisation classes.

\subsection{First polarisation}
\label{subsec:Maxwell-firstpol}

The modes in the first polarisation class are labelled by a pair of
nonnegative integers $(m,n)$, at least one of
which is nonzero, and take the
form
\begin{subequations}
\label{eq:Rindler-em-minusmode}
\begin{align}
A_x &= {k_y} \cos(k_x x) \sin(k_y y) \, g(\eta,\chi)\ ,\\
A_y &= - {k_x} \sin(k_x x) \cos(k_y y) \, g(\eta,\chi)\ ,\\
A_\chi &= 0\ ,
\end{align}
\end{subequations}
where
\begin{equation}
\begin{aligned}
    g(\eta,\chi) &:= \big[I_{-i\Omega}(\kperp \chi_0) I_{i\Omega}(\kperp \chi)\\
      &\hspace{10mm} -I_{i\Omega}(\kperp \chi_0) I_{-i\Omega}(\kperp \chi)\big]e^{-i\Omega \eta}\ ,
  \end{aligned}
\end{equation}
with $k_x = \pi m / L_x$, $k_y = \pi n / L_y$, $\kperp = \sqrt{k_x^2 +
  k_y^2}$ and the eigenfrequencies $\Omega$ for each $(m,n)$ are
determined by the boundary condition that $A_x$ and $A_y$ vanish at
$\chi=\chi_1$. To avoid cluttering the notation, we have left the
modes unnormalised.

To discuss the small acceleration limit, we introduce the coordinates
$(t,\ztilde, x, y)$ by $\eta = h t/L_z$ and $\chi = \chi_0+\ztilde$,
in which the $h\to0$ limit of the metric \eqref{eq:3+1Rindlermetric}
is $ds^2 = - dt^2 + d\ztilde^2 + dx^2 + dy^2$ and the cavity
becomes in this limit static with respect to the Minkowski time
translation Killing vector $\partial_t$ at $0\le\ztilde\le L_z$.
The solutions \eqref{eq:Rindler-em-minusmode} reduce to
\begin{subequations}
\label{eq:Minkowski-em-minusmode}
\begin{align}
A_x &=
{k_y}
\cos(k_x x) \sin(k_y y) \, \tilde{g}(t,\ztilde)
\ ,
\\
A_y &=
- {k_x}
\sin(k_x x) \cos(k_y y) \, \tilde{g}(t,\ztilde)
\ ,
\\
A_{\tilde z} &= 0
\ ,
\end{align}
\end{subequations}
where
\begin{equation}
\tilde{g}(t,\tilde{z}):=\sin(k_z \ztilde) \, e^{-i\sqrt{k_x^2 + k_y^2 + k_z^2} \, t}\ ,
\end{equation}
with $k_z = \pi p / L_z$ with $p=1,2,\dots$.
(The special case of $m=0$ in~\eqref{eq:Minkowski-em-minusmode}
was considered
in~\cite{Ford:2009ci}.)

Comparing \eqref{eq:Rindler-em-minusmode} and
\eqref{eq:Minkowski-em-minusmode} to \eqref{MinkowskiSolutions-all}
and \eqref{RindlerSolutions-function} shows that the modes for
fixed $(m,n)$ are equivalent to the $(1+1)$-dimensional Dirichlet
scalar field discussed in Section \ref{sec:scalar-Dirichlet} with
$\baremass = \kperp$.
The Bogoliubov transformation between an inertial cavity and a
uniformly accelerated cavity can be read off directly from the results
given in Sec.~\ref{sec:scalar-Dirichlet}.

\subsection{Second polarisation}
\label{subsec:Maxwell-secondpol}

The modes in the second polarisation class are labelled by a pair of
positive integers $(m,n)$ and take the form
\begin{subequations}
\label{eq:Rindler-em-plusmode}
\begin{align}
A_x &=
k_x
\cos(k_x x) \sin(k_y y)
\, \chi u(\eta,\chi)
\ ,
\\
A_y &=
k_y
\sin(k_x x) \cos(k_y y)
\, \chi u(\eta,\chi)
\ ,
\\
A_\chi &=
\kperp
\sin(k_x x) \sin(k_y y)
\, \chi v(\eta,\chi)
\ ,
\label{eq:Rindler-em-plusmode-chi}
\end{align}
\end{subequations}
where
\begin{subequations}
\begin{align}
  &\begin{aligned}
    u(\eta,\chi) &:= \big[I'_{-i\Omega}(\kperp \chi_0) I'_{i\Omega}(\kperp \chi)\\
      &\hspace{5mm} - I'_{i\Omega}(\kperp \chi_0) I'_{-i\Omega}(\kperp \chi)\big]e^{-i\Omega \eta}\ ,
  \end{aligned}\\
  &\begin{aligned}
    v(\eta,\chi) &:= \big[I'_{-i\Omega}(\kperp \chi_0) I_{i\Omega}(\kperp \chi)\\
      &\hspace{5mm} - I'_{i\Omega}(\kperp \chi_0) I_{-i\Omega}(\kperp \chi)\big]e^{-i\Omega \eta}\ ,
  \end{aligned}
\end{align}
\end{subequations}
and again $k_x = \pi m / L_x$, $k_y = \pi n / L_y$,
$\kperp = \sqrt{k_x^2 + k_y^2}$ and
the eigenfrequencies $\Omega$ for each $(m,n)$ are
determined by the boundary condition that $A_x$ and $A_y$
vanish at $\chi=\chi_1$. In the small acceleration limit, the
solutions \eqref{eq:Rindler-em-plusmode} reduce to
\begin{subequations}
\label{eq:Minkowski-em-plusmode}
\begin{align}
A_x &=
- k_x k_z
\cos(k_x x) \sin(k_y y)
\, \tilde{u}(t,\ztilde)
\ ,
\\
A_y &=
- k_y k_z
\sin(k_x x) \cos(k_y y)
\, \tilde{u}(t,\ztilde)
\ ,
\\
A_{\tilde z} &=
\kperp^2 \sin(k_x x) \sin(k_y y)
\, \tilde{v}(t,\ztilde)
\ ,
\end{align}
\end{subequations}
where
\begin{subequations}
\begin{align}
  &\begin{aligned}
    \tilde{u}(t,\ztilde) &:= \sin(k_z \ztilde) \, e^{-i\sqrt{k_x^2 + k_y^2 + k_z^2} \, t}\ ,
  \end{aligned}\\
  &\begin{aligned}
    \tilde{v}(t,\ztilde) &:= \cos(k_z \ztilde) \, e^{-i\sqrt{k_x^2 + k_y^2 + k_z^2} \, t}\ ,
  \end{aligned}
\end{align}
\end{subequations}
with $k_z = \pi p / L_z$ with $p=0,1,2,\dots$.

Comparing \eqref{eq:Rindler-em-plusmode} and
\eqref{eq:Minkowski-em-plusmode} to
\eqref{MinkowskiSolutions-function-N}
and
\eqref{RindlerSolutions-function-N}
shows that the eigenfrequencies for fixed
$(m,n)$ are those of the $(1+1)$-dimensional Neumann scalar field
discussed in Section \ref{sec:scalar-Neumann} with $\baremass = \kperp$.
The Bogoliubov transformation between the inertial
cavity and a uniformly accelerated cavity requires a further analysis
because of the contributions from $A_x$ and $A_y$
to the inner product~\eqref{eq:Maxwell-ip}.
The outcome of this analysis is that for fixed $(m,n)$, the
Bogoliubov coefficients are obtained from those of the
$(1+1)$-dimensional Neumann scalar field
of Section \ref{sec:scalar-Neumann} with $\baremass = \kperp$
via the replacement $\mrbeta \to -\mrbeta$.
To linear order in~$h$, the Bogoliubov coefficients can hence
be read off from \eqref{eq:neumann-bogocoeffs-linear} with the replacement
$\mrbeta \to -\mrbeta$.

\subsection{Time-dependent acceleration}
\label{subsec:Maxwell-timedependentacc}

Given the above results about the two polarisation classes, a cavity
with time-dependent acceleration in the $z$ direction can be handled
with the $(1+1)$ scalar field results of Secs.~\ref{sec:scalar-Dirichlet}
and~\ref{sec:scalar-Neumann}. To linear
order in the acceleration, the Bogoliubov coefficient matrices between
initial and final inertial regions are given
by~\eqref{eq:ABhat-int-components}, where $\hat\alpha_{mn}(M)$ and
$\hat\beta_{mn}(M)$ for the first (second) polarisation class are obtained
from the Dirichlet (Neumann) scalar field expressions of Section
\ref{sec:scalar-Dirichlet} (\ref{sec:scalar-Neumann}) with  $\baremass = \kperp$,
with an additional minus sign for the beta-coefficients in the second
polarisation class.

\section{$(1+1)$ massive Dirac spinor}
\label{sec:Diracspinor}

In this section we address a massive Dirac spinor in
$(1+1)$-dimensional Minkowski spacetime, generalising the massless
spinor analysis of \cite{Friis:2011yd} to strictly positive mass.
A~spinor in $(3+1)$-dimensional Minkowski spacetime reduces to
the $(1+1)$-dimensional case by a Fourier decomposition in the
dimensions transverse to the acceleration, with the transverse momenta
making a strictly positive contribution to the effective
$(1+1)$-dimensional mass.

\subsection{Inertial cavity}

In the $(1+1)$-dimensional Minkowski metric~\eqref{eq:1+1Minkowski},
the massive Dirac equation takes the form \cite{srednicki-book}
\begin{align}
i\partial_t {\psi}
=\,\left(-i\alpha_{3}\partial_z + \baremass \beta \right) {\psi}
\ ,
\label{eq:dirac-in-Minkowski}
\end{align}
where the Hermitean matrices $\alpha_3$ and $\beta$ anticommute and
square to the identity.  We assume the mass $\baremass$ to be strictly
positive.  In the present $(1+1)$ setting, we may work with
two-component spinors and introduce a spinor basis $(U_+,U_-)$ that is
orthonormal, in the sense of $U_+^{\dagger} U_- = U_-^{\dagger} U_+
=0$ and $U_+^{\dagger} U_+ = U_-^{\dagger} U_-=1$, and satisfies
\begin{align}
\alpha_3 U_\pm = \pm U_\pm
\ \ , \ \
\beta U_\pm = U_\mp
\ .
\end{align}
An example of an explicit representation would be
$\alpha_3 =
\left(\begin{smallmatrix}
0 & 1 \\
1 & 0
\end{smallmatrix}
\right)$,
$\beta =
\left(\begin{smallmatrix}
1 & 0 \\
0 & -1
\end{smallmatrix}
\right)$,
$U_+ =
\frac{1}{\sqrt{2}}
\left(\begin{smallmatrix}
1 \\
1
\end{smallmatrix}
\right)$
and
$U_- =
\frac{1}{\sqrt{2}}
\left(\begin{smallmatrix}
1 \\
-1
\end{smallmatrix}
\right)$.

We introduce a cavity with walls at $z=z_0$ and $z=z_1 = z_0+L$ as in
Sec.~\ref{sec:scalar-Dirichlet}. The inner product reads
\begin{align}
\bigl({\psi}_{(1)},{\psi}_{(2)}\bigr)
=
\int_{z_0}^{z_1}
dz\,{\psi}_{(1)}^{\dagger}\,{\psi}_{(2)}
\ ,
\label{eq:Minkowski-Dirac-ip}
\end{align}
where we have adopted the convention in which the
fermion inner product is antilinear in the first argument.

We consider boundary conditions that ensure the
vanishing of the probability current independently at each wall,
\begin{align}
{\psi}_{(1)}^{\dagger} \alpha_3 {\psi}_{(2)} \Bigr|_{z=z_0}
=
0
=
{\psi}_{(1)}^{\dagger} \alpha_3 {\psi}_{(2)} \Bigr|_{z=z_1}
\ ,
\label{eq:in-bc}
\end{align}
where ${\psi}_{(1)}$ and ${\psi}_{(2)}$ are any two eigenfunctions of
the Dirac Hamiltonian that appears on the right-hand side
of~\eqref{eq:dirac-in-Minkowski}. An analysis of the deficiency
indices \cite{reebk2,bonneauetal,thaller:dirac} shows that the allowed
boundary conditions are parametrised by a
$\text{U}(1)$ at $z=z_0$ and another $\text{U}(1)$ at $z=z_1$.

Separating the variables, and assuming the eigenvalue $\omega$ of the
Dirac Hamiltonian to satisfy $|\omega| > \baremass$, the linearly
independent solutions to the differential equation
\eqref{eq:dirac-in-Minkowski} can be written as
\begin{subequations}
\label{eq:Mink-spinor-psiplus-psiminus}
\begin{align}
\hspace{-1mm}\psi_{+,k}
&:= \left[\cos(\phi_k) \, U_+ + \sin(\phi_k) \, U_- \right]
e^{ik z-i\omega_k t},
\\
\hspace{-1mm}\psi_{-,k}
&:=
\left[\sin(\phi_k) \, U_+ +  \cos(\phi_k) \, U_- \right]
e^{-ik z-i\omega_k t},
\end{align}
\end{subequations}
where $k \in \BbbR\setminus\{0\}$,
$\omega_k := \sgn(k) \sqrt{\baremass^2+k^2}$
and
$\phi_k = \tfrac12
\arctan(\baremass/k)$.
$\psi_{+,k}$~is a
right-mover and $\psi_{-,k}$ is a left-mover,
and the sign of the frequency is the sign of~$k$.
Imposing \eqref{eq:in-bc} at $z=z_0$ leads to the
linear combination
\begin{subequations}
\label{eq:Mink-spinor-z0bc}
\begin{align}
  &\begin{aligned}
   \hspace{6mm}\mathllap{\psi}&\phantom{:}= \Lambda_{+,k}\ e^{-ik z_0}\,\psi_{+,k}+\Lambda_{-,k}\ e^{ik z_0}\,\psi_{-,k}\ ,
  \end{aligned}\\
  &\begin{aligned}
    \hspace{6mm}\mathllap{\Lambda_{\epsilon,k}}&:= e^{-i\epsilon\pi/4} \cos(\phi_k-\alpha_0)
     - e^{i\epsilon\pi/4} \sin(\phi_k+\alpha_0)\ ,
  \end{aligned}
\end{align}
\end{subequations}
where the parameter $\alpha_0 \in \BbbR \! \mod \pi$ specifies the
boundary condition at $z=z_0$ and $\epsilon\in\{+,-\}$. Imposing
\eqref{eq:in-bc} at $z=z_1$ leads to an expression similar to
\eqref{eq:Mink-spinor-z0bc} with $z_0 \to z_1$ and $\alpha_0 \to
\alpha_1$, and the parameter $\alpha_1 \in \BbbR \! \mod \pi$
specifies the boundary condition at $z=z_1$. For given $\alpha_0$
and~$\alpha_1$, the eigenmodes with $|\omega|>\baremass$ are hence
obtained by imposing both of these boundary conditions, and the
existence of any additional eigenmodes in the range
$|\omega|\le\baremass$ can then be examined using
\eqref{eq:in-bc}~\cite{thaller:dirac}.

From here on we specialise to the MIT bag boundary
condition~\cite{Chodos:1974je,Elizalde:1997hx}, which arises as a
limit when the cavity field is matched to a field of a different mass
in the exterior of the cavity and the exterior mass is taken to
infinity. This is analogous to the way in which the Dirichlet boundary
condition is singled out in nonrelativistic quantum mechanics in the
limit of a potential wall whose height is taken to
infinity~\cite{walton}. In our notation, the MIT bag boundary
condition reads
$(1- i \beta \alpha_3)\psi \bigr|_{z=z_0} = 0
= (1+ i \beta \alpha_3)\psi \bigr|_{z=z_1}$.
This implies $\alpha_0=0$ and
$\alpha_1=\pi/2$.
We find that the normalised eigenfunctions read
\begin{subequations}
\label{eq:Mink-spinor-MITmode}
\begin{align}
\hspace{-1.4mm}\psi_{k}
&=
N_{k} \!
\left(
e^{-i\phi_k} e^{-ik z_0}
\,
\psi_{+,k}
+
i e^{i\phi_k}
e^{ik z_0}
\,
\psi_{-,k}
\right),\\
\hspace{-1.4mm}N_{k}&=\sqrt{\frac{\omega_k^2}
{2L \bigl(\omega_k^2+ (\baremass/L)\bigr)}} \ ,
\end{align}
\end{subequations}
where $k$ takes the discrete
positive and negative values that satisfy the
transcendental equation
\begin{align}
\frac{\tan(k L)}{k L} = - \frac{1}{\baremass L}
\ .
\label{eq:Mink-spinor-EVformula}
\end{align}
We have chosen the phase in
$\psi_{k}$ \eqref{eq:Mink-spinor-MITmode}
so that when $z=z_0$ and
$t=0$, $\psi_{k}$ is a positive multiple of $U_++iU_-$.
The positive and negative eigenfrequencies appear
symmetrically in the spectrum, and all the eigenfrequencies satisfy
$|\omega| >\baremass$.

In the massless limit,
the modes
\eqref{eq:Mink-spinor-MITmode} reduce to
\begin{align}
\hspace{-.5mm}\psi_n &= \frac{\left[
U_+
e^{i\widehat\omega_n (z-z_0)}
+ i
U_-
e^{-i\widehat\omega_n (z-z_0)}
\right]}{\sqrt{2L}}e^{-i\widehat\omega_n \, t},
\label{eq:Mink-spinor-MITmode-massless}
\end{align}
where $\widehat\omega_n := \pi(n + \tfrac12)/L$ with $n\in\BbbZ$.
The positive and negative frequencies appear
symmetrically in the spectrum and there is no zero mode.
Among the massless boundary conditions
classified in~\cite{Friis:2011yd},
\eqref{eq:Mink-spinor-MITmode-massless} is the case $s=1/2$ and
$\theta=\pi/2$.

\subsection{Accelerated cavity}
\label{subsec:fermion-Rindler}

We write the $(1+1)$-dimensional Rindler metric \eqref{eq:1+1Rindlermetric}
as
\begin{align}
ds^2 = - \bigl(e^{\hspace*{1pt}\underline 0}_{\hspace*{1pt}\eta}\bigr)^2 d\eta^2
+ \bigl(e^{\hspace*{1pt}\underline 1}_{\hspace*{1pt}\chi}\bigr)^2 d\chi^2
\ ,
\end{align}
where the nonvanishing components of the co-dyad $e_a^{\underline A}$
are $e_\eta^{\underline 0} = \chi$ and $e_\chi^{\underline 1} = 1$.
The underlined indices are internal Lorentz indices, raised and
lowered with the internal Lorentz metric. The
nonvanishing components of the corresponding dyad $e^a_{\underline A}$ are
\begin{align}
e^\eta_{\underline 0} = 1/\chi
\ , \ \
e^\chi_{\underline 1} = 1
\ .
\label{eq:rightward-dyad}
\end{align}
In the dyad~\eqref{eq:rightward-dyad}, the massive Dirac equation
takes the form
\cite{byd,mcmahonalsingembid06,Langlois:2004fv,Langlois:2005if}
\begin{align}
i\partial_{\eta}{\psi}
=\left(-i \alpha_{3}(\chi\,\partial_{\chi}
+\tfrac{1}{2}) + \baremass \chi \beta \right) \! {\psi}
\ .
\label{eq:dirac-in-Rindler}
\end{align}

We introduce a cavity as in Sec.~\ref{sec:scalar-Dirichlet}, with
walls at $\chi=\chi_0$ and $\chi = \chi_1$ as given by
\eqref{eq:h-introd} with $0<h<2$. The inner product reads
\begin{align}
\bigl({\psi}_{(1)},{\psi}_{(2)}\bigr)
=
\int_{\chi_0}^{\chi_1}
d\chi\,{\psi}_{(1)}^{\dagger}\,{\psi}_{(2)}
\ .
\label{eq:Rindler-Dirac-ip}
\end{align}
We adopt boundary conditions that ensure vanishing of the
probability current through each wall. These boundary conditions read as in
\eqref{eq:in-bc} but with $z\to\chi$.

Separating the variables, we find that the linearly independent
solutions to \eqref{eq:dirac-in-Rindler} are
\begin{subequations}
\label{eq:Rind-spinor-psiplus-psiminus}
\begin{align}
\Psi_{+,\Omega} &:= \bigl[ I_{i\Omega - \frac12}(\baremass \chi)\, U_+
+ i I_{i\Omega + \frac12}(\baremass \chi)\, U_-\bigr]
e^{-i\Omega \eta}
\ ,
\\
\Psi_{-,\Omega} &:= \bigl[ I_{-i\Omega + \frac12}(\baremass \chi)\, U_+
+ i I_{-i\Omega - \frac12}(\baremass \chi)\, U_-\bigr]
e^{-i\Omega \eta}
\ ,
\end{align}
\end{subequations}
where $\Omega\in\BbbR$. The condition of a vanishing probability
current at $\chi=\chi_0$ leads to the linear combination
\begin{equation}
\begin{aligned}
\label{eq:Rind-spinor-chi0bc}
		\psi \!&=\!
		\bigl[C I_{-i\Omega - \frac12}(\baremass \chi_0)-D I_{-i\Omega + \frac12}(\baremass 							 \chi_0)\bigr]
		\Psi_{+,\Omega}
		\\
		& \hspace{3mm}
		+
		\bigl[D I_{i\Omega - \frac12}(\baremass \chi_0)-C I_{i\Omega + \frac12}(\baremass \chi_0)\bigr]
		\Psi_{-,\Omega},
\end{aligned}
\end{equation}
with the coefficients
\begin{subequations}
\begin{align}
C &:=1 + B_0 \tanh(\baremass \chi_0),\\
D &:=B_0 + \tanh(\baremass \chi_0),
\end{align}
\end{subequations}
where the complex number $B_0$ of unit modulus is the parameter that
specifies the boundary condition at $\chi=\chi_0$. The condition of a
vanishing probability current at $\chi=\chi_1$ leads to a similar
expression with $\chi_0 \to \chi_1$ and $B_0 \to B_1$, where the
complex number $B_1$ of unit modulus is the parameter that specifies
the boundary condition at $\chi=\chi_1$.

We again specialise to the MIT bag boundary condition, which now reads
$(1- i \beta \alpha_3)\psi \bigr|_{\chi=\chi_0} = 0 = (1+ i \beta
\alpha_3)\psi \bigr|_{\chi=\chi_1}$. This implies $B_0=1$ and
$B_1=-1$. The normalised eigenfunctions read
\begin{equation}
\label{eq:Rind-spinor-MITmode}
\begin{aligned}
\hspace{-1mm}\psi_\Omega &= N_\Omega \Big\{\!
\bigl[
I_{-i\Omega - \frac12}(\baremass \chi_0)
\!-\! I_{-i\Omega + \frac12}(\baremass \chi_0)
\bigr]\Psi_{+,\Omega} \\
& \hspace{5mm} + \bigl[
I_{i\Omega - \frac12}(\baremass \chi_0)
\!-\! I_{i\Omega + \frac12}(\baremass \chi_0)
\bigr]\Psi_{-,\Omega}
\Big\},
\end{aligned}
\end{equation}
where $\Omega$ takes the discrete real values that satisfy
\begin{equation}
\label{eq:Rind-spinor-EVformula}
P_{-}P_{+}+\overline{P}_{-}\overline{P}_{+}=0,
\end{equation}
where
\begin{subequations}
\begin{align}
P_{-}&:=I_{-i\Omega - \frac12}(\baremass \chi_0)-I_{-i\Omega + \frac12}(\baremass \chi_0),\\
P_{+}&:=I_{-i\Omega - \frac12}(\baremass \chi_1)+I_{-i\Omega + \frac12}(\baremass \chi_1),
\end{align}
\end{subequations}
and $N_\Omega$ is a normalisation constant.
As \eqref{eq:Rind-spinor-EVformula} is invariant under $\Omega \to
-\Omega$, the positive and negative eigenfrequencies appear
symmetrically in the spectrum.

In the massless limit, the modes \eqref{eq:Rind-spinor-MITmode} reduce
to those given \cite{Friis:2011yd}
with $s=1/2$ and $\theta=\pi/2$. The symmetry between the positive and
negative frequencies hence persists in the massless limit, and the
massless field has no zero mode.\\

\subsection{Matching}

We match an inertial cavity at $t\le0$ to an accelerated cavity at
$\eta\ge0$ across the hypersurface $t=0$ as in
subsection~\ref{subsec:dirichlet-matching}.
We assume to begin with that the
acceleration is towards increasing~$z$,
so that the Minkowski and Rindler
coordinates are related by~\eqref{eq:Rindlertransformation}.
It follows that the time and space orientations of the dyad
\eqref{eq:rightward-dyad}
agree with those of the Minkowski coordinates $(t,z)$.
We may hence write the Bogoliubov transformation from the
Minkowski modes
\eqref{eq:Mink-spinor-MITmode} to the Rindler
modes \eqref{eq:Rind-spinor-MITmode} as
\begin{align}
\label{eq:spinor-bogo}
\Psi_\Omega
= \sum_{k} \mrA_{\Omega k} \, \psi_k
\ ,
\end{align}
where the Bogoliubov coefficient matrix
$\mrA = \left(\mrA_{\Omega k}\right)$
is given by
\begin{align}
\mrA_{\Omega k}
= \bigl(\psi_{k}, \Psi_\Omega \bigr)
\ ,
\label{eq:spinor-A-ipformulas}
\end{align}
and the inner product
in \eqref{eq:spinor-A-ipformulas}
is evaluated on the surface $t=0$.
By the orthonormality of the Minkowski modes
and the orthonormality of
the Rindler modes, $\mrA$ is unitary.

At small~$h$, matching the Rindler modes \eqref{eq:Rind-spinor-MITmode}
with the Minkowski modes \eqref{eq:Mink-spinor-MITmode} shows that
the leading term in $\Omega$ must be proportional to~$1/h$.
The order of the Bessel functions has hence a phase that
approaches $\pm\pi/2$ as $h\to0$,
which is a regime of subtlety in the uniform asymptotic expansions
of Bessel functions for
large complex order~\cite{olver-as-bessel}.
We therefore expand the Rindler modes in $h$
starting directly from the Bessel differential equation that leads
to the solutions~\eqref{eq:Rind-spinor-psiplus-psiminus}, writing
$\Omega = \Omega_{-1}/h + \Omega_0 + \Omega_{1}h + \cdots$ and
$\chi = (L/h)(1 + hv)$, where the new dimensionless spatial coordinate
$v$ has been chosen so that $\chi=\chi_0$ at
$v=-\frac12$ and $\chi=\chi_1$ at $v=\frac12$.
We find that the eigenvalues of $\Omega$ have the form
\begin{align}
\Omega_k = L h^{-1} \omega_k \left(1 + O(h^2) \right)
\ ,
\end{align}
where the index $k$ takes the discrete
positive and negative values that
satisfy~\eqref{eq:Mink-spinor-EVformula}.
Choosing the phase of $N_\Omega$ so that the phases of the
Rindler modes \eqref{eq:Rind-spinor-MITmode}
match those of the Minkowski modes \eqref{eq:Mink-spinor-MITmode}
at $t=0$, we find
that the Bogoliubov coefficients to linear order in $h$ read

\begin{widetext}
\begin{subequations}
\label{eq:Dirac-bogocoeffs-linear}
\vspace*{-4mm}
\begin{align}
\mrA_{\Omega_k k}  &= 1 + O(h^2)
\ ,
\\[1ex]
\mrA_{\Omega_k l}  &=
\frac{2\bigl({(-1)}^{n_k+n_l}-1 \bigr)
|k l| C_k^2 C_l^2 (C_k+C_l)
(C_kC_l + \baremass^2)}
{\sqrt{L^2 \omega_k^2 + \baremass L}
\sqrt{L^2 \omega_l^2 + \baremass L}
\,
{(C_k-C_l)}^3
{(C_kC_l - \baremass^2)}^3
}
\, h
+ O(h^2)
\ \ \ \hbox{for $k\ne l$},
\end{align}
\end{subequations}
\end{widetext}
where $C_k := \omega_k+k$ and $n_k\in\BbbZ$ is such that the map
$k\mapsto n_k$ indexes the consecutive solutions to
\eqref{eq:Rind-spinor-psiplus-psiminus} by consecutive integers. As a
consistency check, we note that
the order $h$ term in \eqref{eq:Dirac-bogocoeffs-linear} is
anti-Hermitian, as it must be by unitarity of~$\mrA$.
As another consistency check, we note that in the massless limit
\eqref{eq:Dirac-bogocoeffs-linear} reduces to the expressions given
in \cite{Friis:2011yd} with $s=1/2$.

Suppose then that the acceleration is towards decreasing~$z$. We
introduce the leftward Rindler coordinates $(\tilde\eta,\tilde\chi)$
by \eqref{eq:leftRindlertransformation} and the compatible dyad
$\tilde e^a_{\underline A}$ whose nonvanishing components are
\begin{align}
\tilde e^{\tilde\eta}_{\underline 0} = 1/\tilde\chi
\ , \ \
\tilde e^{\tilde\chi}_{\underline 1} = - 1
\ .
\label{eq:leftward-dyad}
\end{align}
The time and space orientations of this dyad agree with those
of the Minkowski coordinates $(t,z)$.
Because of the minus sign in~\eqref{eq:leftward-dyad},
the Dirac equation reads as in \eqref{eq:dirac-in-Rindler}
but with tildes on the coordinates and with the replacement
$\alpha_{3} \to - \alpha_{3}$.
It follows that the separation of
variables proceeds as in subsection \ref{subsec:fermion-Rindler} but
with $U_+$ and $U_-$ interchanged.

With the MIT bag boundary conditions,
it is hence seen from \eqref{eq:Mink-spinor-psiplus-psiminus},
\eqref{eq:Mink-spinor-MITmode},
and \eqref{eq:Mink-spinor-EVformula}
that the leftward acceleration Bogoliubov coefficients are obtained
from the rightward ones by keeping $h$ positive but
inserting in $\mrA_{\Omega_k l}$ the phase factors
$\exp[i(2\phi_k - k L)]\exp[i(2\phi_l - l L)] = {(-1)}^{n_k+n_l}$.
It follows that the formulas \eqref{eq:Dirac-bogocoeffs-linear}
cover both directions of acceleration provided positive
(respectively, negative)
values of $h$ are taken to indicate acceleration towards
increasing (decreasing)~$z$.
We have verified that the same holds also when
the $h^2$ terms are included
in~\eqref{eq:Dirac-bogocoeffs-linear}.

\subsection{Time-dependent acceleration}

Time-dependent acceleration for the spinor field can be handled as for
the bosonic fields. For acceleration that is piecewise constant in
time, the above Minkowski-to-Rindler transformation and its inverse
can be used to compose inertial and uniformly accelerated
segments. For motion in which the acceleration is not necessarily
piecewise constant, we can pass to the limit: proceeding as
in~\cite{Bruschi:2012pd}, we find that to linear order in the
acceleration the Bogoliubov coefficient matrix between an initial
inertial region at $\tau\le\tau_0$ and a final inertial region at
$\tau\ge\tau_f$ reads
\begin{subequations}
\label{eq:Dirac-smoothbogos}
\begin{align}
  &\begin{aligned}
    \smoothA_{\Omega_k k} &= e^{i{\omega}_k (\tau_f-\tau_0)} \ ,
  \end{aligned}\\
  &\begin{aligned}
    \smoothA_{\Omega_k l} &= i L (\omega_k-\omega_l) \hat A_{\Omega_k l}(M) \, e^{i{\omega}_k (\tau_f-\tau_0)}\\
	&\times\int_{\tau_0}^{\tau_f}e^{-i({\omega}_k - {\omega}_l) (\tau-\tau_0)} \, a(\tau) \, d\tau \\
	&\hspace{35mm}\hbox{(for $k\ne l$)},
  \end{aligned}
\end{align}
\end{subequations}
where $\hat A_{\Omega_k l}(M)$ denotes the coefficient of $h$ in the
expansion~\eqref{eq:Dirac-bogocoeffs-linear}, and we have indicated
explicitly the dependence of this coefficient on $\baremass$ and $L$
through the dimensionless combination $M := \baremass L$.

\subsection{Applications in quantum information}

Let us now consider the implications of the MIT bag boundary
conditions and the extension to massive Dirac spinors for quantum
information tasks. Since the aim of this paper lies in the analysis of
different boundary conditions and masses of the field excitations we
are not introducing a detailed description of quantum information
theory with modes of quantum fields. For a recent investigation of the
description and issues of fermionic density operator constructions for
quantum information purposes see~\cite{FriisLeeBruschi2013}.

Instead, we shall discuss the direct consequences on some quantities
of interest that can be expressed directly in terms of the cavity
Bogoliubov coefficients. In particular, the results of the present
paper allow us to extend the validity of the expressions obtained
in~\cite{Friis:2011yd,Friis:2012tb,Friis:2012ki} to massive
$(1+1)$-dimensional spinor fields. Two distinct cases of interest are
affected: degradation effects, reducing the amount of entanglement
that is shared between two modes situated in different
cavities~\cite{Friis:2011yd}, as opposed to entanglement generation
between modes within a single cavity~\cite{Friis:2012tb,Friis:2012ki}.

For entanglement degradation effects, the inclusion of mass and
transverse momenta and the choice of boundary conditions result in
quantitative changes of the amount of decoherence. Qualitatively,
nonzero effective mass removes the periodicity in the duration of
individual segments of motion for travel scenarios of piecewise
constant acceleration. For a massless field in $(1+1)$
dimensions, on the other hand, the Bogoliubov coefficients are
periodic in the duration of such segments~\cite{Friis:2011yd}.

\begin{figure}[t]
(a)\includegraphics[width=0.45\textwidth]{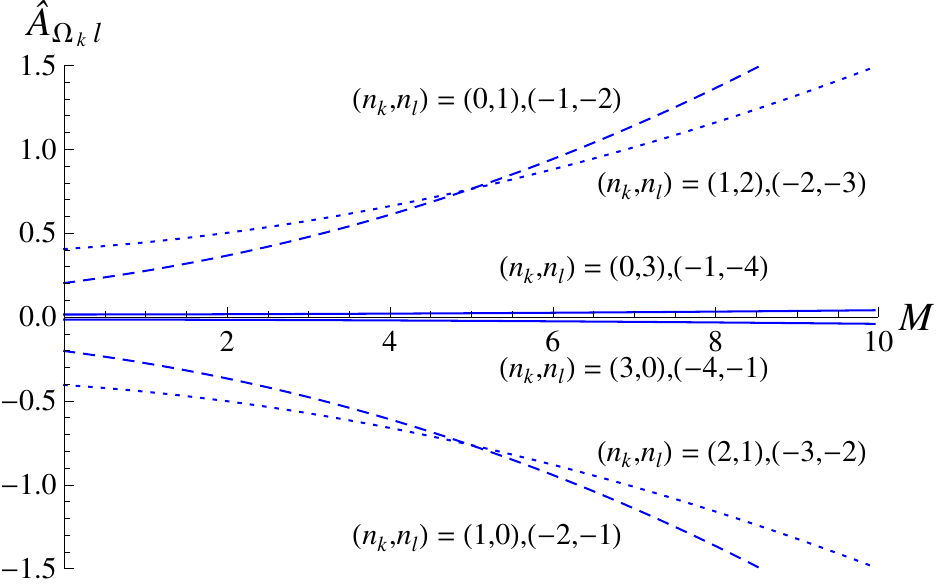}
(b)\includegraphics[width=0.45\textwidth]{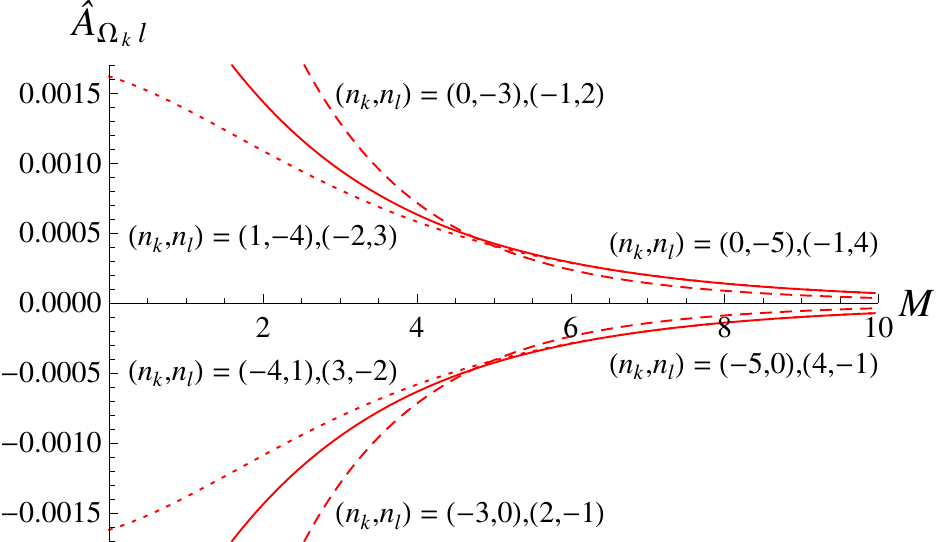}
\caption{\label{fig:alphas and betas}Behaviour of the Bogoliubov coefficients for the massive $(1+1)$-dimensional Dirac field with MIT bag boundary conditions for increasing mass. The coefficient of $h$ in \eqref{eq:Dirac-bogocoeffs-linear}, denoted by $\hat A_{\Omega_k l}(M)$, is plotted against the dimensionless combination $M := \baremass L$. Figure \ref{fig:alphas and betas}~(a) shows a selection of Bogoliubov coefficients that relate modes with the same sign of the frequency ($\alpha$-type coefficients):
the map $k\mapsto n_k$ has been chosen so that $n_{k}\geq0$ labels the positive frequency solutions and $n_{k}<0$ labels the negative frequency solutions. These mode-mixing coefficients are proportional to $M^{2}$ as $M\rightarrow\infty\,$. Figure \ref{fig:alphas and betas}~(b) shows a selection of Bogoliubov coefficients that relate positive frequency modes with negative frequency modes ($\beta$-type coefficients). These particle creation coefficients are proportional to $M^{-6}$ as $M\rightarrow\infty\,$.}
\end{figure}

In scenarios where entanglement generation between two or more modes
in a single cavity is considered, the leading-order effects are
determined by the coefficients of $h$ in
\eqref{eq:Dirac-bogocoeffs-linear}~\cite{Friis:2011yd}. As noted
above, we denote these coefficients by~$\hat A_{\Omega_k l}(M)$,
indicating explicitly their dependence on $\baremass$ and $L$ through
the dimensionless combination $M := \baremass L$.  Selected plots are
shown in Fig.~\ref{fig:alphas and betas}.  In the limit
$M\to\infty$, it can be shown from \eqref{eq:Dirac-bogocoeffs-linear}
that the mode-mixing coefficients increase proportionally to~$M^{2}$
[Figure \ref{fig:alphas and betas}~(a)], while the particle-creation
coefficients decrease proportionally to~$M^{-6}$ [Figure \ref{fig:alphas
  and betas}~(b)].  The relevance of this behaviour becomes apparent
when we consider initial pure states of two modes labelled by $n_{k}$
and $n_{k^{\prime}}$, respectively.  The coefficient $\hat A_{\Omega_k
  k^{\prime}}$ is directly related to the entanglement that is created
between these modes due to the cavity motion~\cite{Friis:2012tb}. The
qualitatively different dependence on the field mass for mode-mixing
and particle-creation Bogoliubov coefficients indicates that nonzero
mass enhances entanglement generation between modes of equal charge,
while the effect is suppressed between modes of opposite charge.

Finally, it is also of interest to reconsider the massless limit. As
noted before, the coefficients for the MIT bag boundary condition
reduce to the case $s=\tfrac{1}{2}$ (rather than $s=0$) discussed
in~\cite{Friis:2011yd}. Since this choice removes the zero mode from
the spectrum, the resulting Bogoliubov coefficients allow for a
violation of the Clauser-Horne-Shimony-Holt inequality
\cite{ClauserHorneShimonyHolt1969,HorodeckiRPM1995} by the
entanglement generated from the initial vacuum state.

\section{Unitarity of evolution}
\label{sec:unitarity}

In this section we address the unitary implementability of the cavity
field's time evolution, for both smoothly varying and sharply varying
accelerations.  We treat the boson fields and the spinor field in turn.

\subsection{Bosons}

Recall that a Bogoliubov transformation for a real bosonic field, with
the coefficient matrices written in our notation as $\alpha =
(\alpha_{ij})$ and $\beta = (\beta_{ij})$~\cite{byd}, is implementable
as a unitary transformation iff the matrix $\alpha^{-1}\beta$ is
Hilbert-Schmidt, $\sum_{ij} {|(\alpha^{-1}\beta)_{ij}|}^2 <
\infty$~\cite{shale-unitarity,shale-stinespring,honegger-rieckers,labonte}.


We start with the $(1+1)$-dimensional scalar field of Secs.~\ref{sec:scalar-Dirichlet} and~\ref{sec:scalar-Neumann}, and with the
Bogoliubov transformation from the inertial segment to the uniformly
accelerated segment.  While the perturbative small acceleration
expansions of the Bogoliubov coefficients in
\eqref{eq:dirichlet-bogocoeffs-linear} and
\eqref{eq:neumann-bogocoeffs-linear} are not uniform in the
mode numbers, we may nevertheless examine the unitarity of the
dynamics perturbatively in~$h$. To leading order in~$h$, this reduces
to considering the linear terms in
\eqref{eq:dirichlet-bogocoeffs-linear}
and~\eqref{eq:neumann-bogocoeffs-linear}, and to this order the
Hilbert-Schmidt condition for $\mralpha^{-1}\mrbeta$ is
equivalent to the Hilbert-Schmidt condition for~$\mrbeta$.

In the notation established in Secs.~\ref{sec:scalar-Dirichlet}
and~\ref{sec:scalar-Neumann}, we denote the coefficient of $h$ in the
expansion of $\mrbeta_{mn}$ in \eqref{eq:dirichlet-betacoeffs-linear}
or \eqref{eq:neumann-betacoeffs-linear} by~$\hat\beta_{mn}(M)$,
continuing to suppress the distinction between Dirichlet and Neumann,
but indicating explicitly the dependence on $\baremass$ and $L$
through the dimensionless combination $M := \baremass L$.  Elementary
estimates show that the function $F(M) := \sum_{mn}
\bigl|\hat\beta_{mn}(M)\bigr|^2$ is finite for all values of~$M$. The
field evolution in the sharp transition from the inertial segment to
the uniformly accelerated segment is hence perturbatively unitary to
linear order in~$h$.

Suppose then that the acceleration varies smoothly between an initial
inertial region and a final inertial region, so that to linear order
in the acceleration the Bogoliubov coefficients are given
by~\eqref{eq:ABhat-int-components}, where $\hat\alpha_{mn}(M)$ and
$\hat\beta_{mn}(M)$ are the coefficients of $h$ in the expansions of
$\mralpha_{mn}$ and $\mrbeta_{mn}$
\eqref{eq:dirichlet-bogocoeffs-linear}
or~\eqref{eq:neumann-bogocoeffs-linear}. As the Fourier transform of a
smooth function of compact support falls off at infinity faster than
any power, \eqref{eq:Bhat-int-components} shows that
$\smoothbeta_{mn}$ is bounded in absolute value by
$\bigl|\hat\beta_{mn}(M) f(\omega_m+\omega_n)\bigr|$, where $f$ is a
function that falls off at infinity faster than any power. The sum
$\sum_{mn} \bigl|\smoothbeta_{mn}\bigr|^2$ is hence finite. We
conclude that the evolution is perturbatively unitary to linear order
in the acceleration.

Consider then a rectangular cavity in a higher-dimensional spacetime,
with acceleration in one of its principal directions.  By Fourier
decomposition in the transverse dimensions, the Bogoliubov
transformation reduces to that of the $(1+1)$-dimensional cavity for
each set of the transverse quantum numbers, with the transverse
momenta contributing to the effective $(1+1)$-dimensional mass.  The
trace in the Hilbert-Schmidt norm includes now also a sum over the
transverse quantum numbers.  For the sharp evolution from inertial
motion to uniform acceleration, the criterion of leading-order
perturbative unitarity is hence the finiteness of the sum
$\sum_{\mathbf{k}_\perp} F(L\sqrt{\mu_0^2 + \mathbf{k}^2_\perp})$,
where $\mu_0$ is the genuine mass and $\mathbf{k}_\perp$ are the
quantised transverse momenta. It follows from the estimates given in
the Appendix that this criterion is satisfied in $(2+1)$ dimensions
but not in $(3+1)$ or higher. The perturbative unitarity of the
dynamics hence fails in $(3+1)$ spacetime dimensions and above when
the onset of the acceleration is sharp.  When the acceleration changes
smoothly, by contrast, the rapid falloff of $\sum_{mn}
\bigl|\smoothbeta_{mn}\bigr|^2$ guarantees that the evolution is
unitary in any spacetime dimension.

Finally, as the Maxwell field in a $(3+1)$-dimensional cavity
decomposes into Dirichlet-type polarisation modes and Neumann-type
polarisation modes, the results about the perturbative unitarity of
the time evolution follow directly from those for the scalar
field. Unitarity holds when the acceleration changes smoothly but
fails when the acceleration onset is sharp.

\subsection{Fermions}

For a fermionic field, a Bogoliubov transformation is unitarily
implementable if the two blocks of the Bogoliubov transformation
matrix that relate positive frequencies to negative frequencies are
Hilbert-Schmidt
\cite{thaller:dirac,shale-unitarity,shale-stinespring,honegger-rieckers,labonte}.
We consider this condition in our system perturbatively in the
acceleration, to the leading order.

Consider first the $(1+1)$-dimensional Dirac field of Section \ref{sec:Diracspinor}
and the Bogoliubov transformation from the inertial segment to the uniformly
accelerated segment. Recall that we denote the
coefficient of $h$ in \eqref{eq:Dirac-bogocoeffs-linear}
by $\hat A_{\Omega_k l}(M)$, indicating explicitly
the dependence on $\baremass$ and $L$
through the dimensionless combination $M := \baremass L$.
The condition of unitarily implementable evolution for given $M$ is then that
$G(M) := \sum_{k>0>l} \bigl|\hat A_{\Omega_k l}(M)\bigr|^2$,
or equivalently $G(M) := \sum_{l>0>k} \bigl|\hat A_{\Omega_k l}(M)\bigr|^2$,
is finite. Elementary estimates show that this condition
holds for all~$M$.

When the acceleration varies smoothly in time,
the Bogoliubov coefficient matrix between an initial inertial region
and a final inertial region is given by~\eqref{eq:Dirac-smoothbogos}.
The rapid falloff of the Fourier transform guarantees that
the unitarity condition is satisfied.

Consider then a rectangular cavity in a higher-dimensional spacetime,
with acceleration in one of its principal directions. Proceeding as
for the scalar field, and using the large $M$ behaviour of $G(M)$
established in the Appendix, we find that the situation is as for the
scalar field: unitarity holds for smoothly varying acceleration in any
spacetime dimension but fails for sharply varying acceleration in
spacetime dimension $(3+1)$ and higher.

\section{Conclusions}
\label{sec:conclusions}

In this paper we have investigated scalar, spinor, and photon fields in
a cavity that is accelerated in Minkowki spacetime.
The cavity was assumed mechanically rigid,
and we worked within a recently introduced perturbative
formalism \cite{Bruschi:2011ug} that assumes accelerations to remain small
compared with the inverse linear dimensions of the cavity
but allows the velocities, travel times, and travel distances to be arbitrary,
and in particular includes the regime where the velocities are relativistic.
We extended previous scalar field analyses to cover
both Dirichlet and Neumann boundary conditions,
and we showed that a photon field in
$(3+1)$ spacetime dimensions with perfect conductor
boundary conditions decomposes into Dirchlet-type
and Neumann-type polarisation modes.
For a Dirac spinor, we extended previous work on $(1+1)$-dimensional
massless spinors to a strictly positive $(1+1)$-dimensional mass:
this is necessary to handle a cavity in dimensions higher than $(1+1)$,
where the dimensions transverse to the acceleration
give a strictly positive contribution to the effective $(1+1)$-dimensional mass.
We also presented the spinor field time evolution formulas for acceleration
with arbitrary time dependence,
in parallel with the scalar field formulas given in~\cite{Bruschi:2012pd}.
We discussed briefly the consequences of the nonvanishing $(1+1)$-dimensional mass for
quantum information tasks with Dirac fermions, noting that the mass
and the absence of a zero mode
can enhance both entanglement degradation and generation effects.

Finally, we considered whether particle creation
in the cavity could become strong enough to prevent
the time evolution of the quantum field from being
implementable as a unitary transformation in the Fock space.
Working to linear order in the acceleration,
we found the evolution to be unitary when the acceleration
varies smoothly in time.
In the limit of discontinously varying accelerations
the evolution remains unitary in spacetime dimensions
$(1+1)$ and $(2+1)$ but becomes nonunitary in spacetime
dimensions $(3+1)$ and higher.

While the focus of this paper was theoretical,
we shall finish by recalling two experimental
situations for which our results are relevant.

First, traditional proposals to observe acceleration
effects in the laboratory use
photons~\cite{moore-dyncas,Reynaud1,Dodonov:advchemphys,Dodonov:2010zza},
and success in observing the generated
photons has been recently
reported in an experiment
where acceleration is simulated by
SQUID circuits~\cite{wilson-etal}.
Our results confirm that the small acceleration
cavity formalism that was introduced in
\cite{Bruschi:2011ug} for a scalar field adapts in a straightforward
way to the electromagnetic field. It follows in particular that
the experimental scenario of mode mixing
in a cavity accelerated on a desktop,
proposed and analysed for a scalar field in~\cite{Bruschi:2012pd},
does apply to photons captured in the cavity.

Second, it has been proposed that acceleration effects for fermions
can be simulated in solid state analogue
systems~\cite{boada:dirac,ZhangWangZhu2012,Iorio2012}. Our work
provides the theoretical framework for analysing such acceleration
effects with cavitylike boundary conditions whenever the fermion
field has a mass and/or dimensions transverse to the acceleration.
\vspace*{2mm}

\section*{Acknowledgements}
We thank
Pablo Barberis-Blostein,
David Bruschi,
Chris Fewster,
Ivette Fuentes,
Carlos Sab\'in,
and
especially
Elizabeth Winstanley
for helpful discussions and correspondence.
N.F. acknowledges support from EPSRC
(CAF Grant No.\ EP/G00496X/2 to Ivette Fuentes).
J.L. was supported in part by
STFC
(Theory Consolidated Grant No.~ST/J000388/1).



\begin{widetext}

\begin{appendix}

\section{Asymptotics of Bogoliubov coefficient sums}

In this Appendix we establish the asymptotic large $M$
behaviour of the functions $F(M)$ and $G(M)$
defined in Sec.~\ref{sec:unitarity}.

Recall that $F(M) := \sum_{mn} \bigl|\hat\beta_{mn}(M)\bigr|^2$, where
$\hat\beta_{mn}(M)$ is the coefficient of $h$ in the expansion of
$\mrbeta_{mn}$ in \eqref{eq:dirichlet-betacoeffs-linear} or
\eqref{eq:neumann-betacoeffs-linear}, where the notation suppresses
the distinction between the Dirichlet and Neumann boundary conditions
but indicates explicitly the dependence on $\baremass$ and $L$ through
the dimensionless combination $M := \baremass L>0$.  Recall similarly
that $G(M) := \sum_{k>0>l} \bigl|\hat A_{\Omega_k l}(M)\bigr|^2$, or
equivalently $G(M) := \sum_{l>0>k} \bigl|\hat A_{\Omega_k
  l}(M)\bigr|^2$, where $\hat A_{\Omega_k l}(M)$ is the coefficient of
$h$ in~\eqref{eq:Dirac-bogocoeffs-linear}.

Elementary estimates show that $F(M)$ and $G(M)$ are finite for
all~$M$. At $M\to\infty$, we find
\begin{subequations}
\begin{align}
  &\begin{aligned}
    M^2 F(M) \to
\begin{cases}
\displaystyle
\frac{2}{\pi^2}
\int
\limits_{\substack{x>0\\y>0}}
\frac{x^2 y^2\, dx \, dy}{\sqrt{1+x^2}
\sqrt{1+y^2} \, \bigl(\sqrt{1+x^2} + \sqrt{1+y^2}\,\bigr)^6}
\ = \frac{1}{90 \pi^2}
\ ,
&
\text{(Dirichlet)}
\\[5ex]
\displaystyle
\frac{2}{\pi^2}
\int
\limits_{\substack{x>0\\y>0}}
\frac{\bigl(\sqrt{1+x^2}
\sqrt{1+y^2} \, +1 \bigr)^2   \, dx \, dy}
{\sqrt{1+x^2} \sqrt{1+y^2} \, \bigl(\sqrt{1+x^2} + \sqrt{1+y^2}\,\bigr)^6}
\ = \frac{11}{90 \pi^2}
\ ,
&
\text{(Neumann)}
\end{cases}
\label{eq:bos-betasum-asymptotics}
  \end{aligned}\\[2mm]
  &\begin{aligned}
    M^2 G(M) &
\to \frac{8}{\pi^2}
\mathop{\mathlarger{\int}}
\limits_{\substack{x>0\\y>0}}
\frac{\bigl(\sqrt{1+x^2} + x - \sqrt{1+y^2} -y\bigr)^2
\bigl[\bigl(\sqrt{1+x^2}+x\bigr) \bigl(\sqrt{1+y^2}+y\bigr) - 1\bigr]^2}
{
\bigl(\sqrt{1+x^2} + x + \sqrt{1+y^2} + y\bigr)^6
\bigl[\bigl(\sqrt{1+x^2}+x\bigr) \bigl(\sqrt{1+y^2}+y\bigr) + 1\bigr]^6
}
\\[1ex]
&
\hspace{10ex}
\times
\frac{x^2 y^2 \, \bigl(\sqrt{1+x^2} + x\bigr)^4 \bigl(\sqrt{1+y^2} + y\bigr)^4}
{(1+x^2) (1+y^2)}
\>
dx\,dy \ ,
\\[1ex]
&
\hspace{3ex}
= \frac{7}{45 \pi^2} - \frac{1}{64}
\ ,
\label{eq:fer-betasum-asymptotics}
  \end{aligned}
\end{align}
\end{subequations}
where the integral expressions ensue
by regarding the sum as a Riemann sum: in
\eqref{eq:bos-betasum-asymptotics} we have set
$x = (\pi/M)m$ and $y = (\pi/M)n$,
and in \eqref{eq:fer-betasum-asymptotics}
we have set
$x = |k|/\baremass$ and $y = |l|/\baremass$.
The integrals can be evaluated by the substitution
$x = \tfrac12(u-u^{-1})$, $y = \tfrac12(v-v^{-1})$.

\end{appendix}

\end{widetext}

\end{document}